\documentclass[journal]{IEEEtran}
\usepackage[pdftex]{graphicx}
\DeclareGraphicsExtensions{.pdf,.jpeg,.png}
\usepackage[caption=false,font=footnotesize]{subfig}

\begin{document}
%
\title{Visible Iris Area as a Quality Metric for Reliable Iris Recognition Under Pupil Dilation and Eyelid Occlusion}
%
%
%

\author{Jack~Pessaud\IEEEauthorrefmark{1}, 
        Eric~Moran\IEEEauthorrefmark{1}, 
        John~Nguyen\IEEEauthorrefmark{1}, 
        and~Joel~Palko\IEEEauthorrefmark{1}%
\thanks{\IEEEauthorrefmark{1} West Virginia University Eye Institute, Morgantown, WV 26506 USA}%
}

\maketitle

\begin{abstract}
With the increasing adoption of iris recognition systems and the expansion of large-scale enrollment databases, there is a growing need to efficiently assess iris image quality at the time of acquisition, particularly to model user non-compliance in real time. Image quality may degrade due to eyelid occlusion or pupil dilation. Although previous studies have shown that occlusion and changes in the pupil-to-iris ratio negatively impact recognition performance, these investigations were typically limited by small sample sizes and did not examine the combined effects of eyelid and pupil variations. In this study, we analyze both dilation and eyelid occlusion using a large dataset of 555 distinct irises and demonstrate a strong correlation between probe image visible iris area and the Hamming distance of iris code pairs. These results suggest that visible iris area is a robust indicator of probe image quality and could be efficiently incorporated into the iris acquisition process to improve confidence in match predictions.
\end{abstract}


%
\IEEEpeerreviewmaketitle

\section{Introduction}

\IEEEPARstart{I}{ris} recognition has become an increasingly prominent method of biometric identification over the past two decades \cite{nguyenDeepLearningIris2024}. India’s national biometric ID system, Aadhaar, has enrolled more than one billion individuals for applications in national security, identity verification, and access to government services such as banking, subsidies, and healthcare \cite{raoAadhaarGoverningBiometrics2019}. Dubai International Airport employs iris recognition for automated border clearance, performing an estimated 2.7 billion comparisons daily \cite{daugmanIrisRecognitionBordercrossing}. The Canadian Border Services Agency’s NEXUS program has enrolled more than 1.3 million irises for expedited border crossings \cite{ortizExploratoryAnalysisOperational2015}, while the World Foundation, steward of the Worldcoin project, has reportedly registered over 15 million individuals in its iris database \cite{worldfoundationIntroducingWorldID2025}.

Irises are valued for biometric identification because of their intrinsic uniqueness and the relative ease of image acquisition compared with other biometric modalities such as DNA or fingerprints \cite{daugmanProbingUniquenessRandomness2006}. However, several factors can reduce recognition accuracy. Iris size can vary with pupil dilation, and partial occlusion from eyelids or eyelashes can obscure features necessary for reliable segmentation. Environmental conditions such as illumination or gaze angle can also degrade image quality and introduce geometric distortions \cite{celikPupilDilationSynthetic2016}. These sources of variability can substantially impact iris matching performance.

\section{Background}
\subsection {General Methods in Iris Recognition}
John Daugman pioneered modern iris recognition through a statistical framework that leverages between-person variability in iris texture \cite{daugmanImportanceBeingRandom2003}. His method, which remains the foundation of most current iris recognition systems, relies critically on accurate segmentation of the iris region to isolate it from surrounding structures such as the pupil, sclera, and eyelids.

Following segmentation, the iris is normalized from Cartesian to polar coordinates to account for geometric variation due to pupil dilation and imaging angle. A Gabor filter is then applied to extract local spatial frequency information, producing complex-valued phase responses with real and imaginary components. These phase responses are quantized into binary representations, forming a compact iris code that encodes the unique texture pattern of an individual’s iris \cite{daugmanImportanceBeingRandom2003}.

Two iris codes can be compared by computing their Hamming distance (HD), which measures the fraction of corresponding bits that disagree. A sufficiently low HD indicates statistical similarity beyond random probability, implying a genuine match. The choice of HD threshold, below which two irises are considered identical, is therefore central to the system’s performance \cite{daugmanBiometricDecisionLandscapes2000}.

The optimal threshold can be estimated by modeling the distributions of HD values for matching and nonmatching pairs. Ideally, these distributions have small variances and well-separated means, producing minimal overlap between genuine and impostor comparisons. The degree of separation can be quantified using the decidability index ($d'$), defined as

\begin{equation}
d' = \frac{|\mu_1 - \mu_2|}{\sqrt{0.5(\sigma_1^2 + \sigma_2^2)}}
\label{decEq}
\end{equation}

where $\mu_1$ and $\mu_2$ are the means and $\sigma_1$, $\sigma_2$ are the standard deviations of the genuine and impostor distributions, respectively \cite{daugmanBiometricDecisionLandscapes2000}. A larger $d'$ indicates greater separability and a more reliable decision environment. Therefore, one key interest in iris recognition is finding the features that most impact the decision environment by either increasing or decreasing the decidability index, typically through having a strong correlation with HD \cite{daugmanImportanceBeingRandom2003}. 

System performance is also characterized by the overlap between the two distributions, typically reported through the false match rate (FMR), defined as the percentage of impostor pairs incorrectly accepted as matches, and the false nonmatch rate (FNMR), defined as the percentage of genuine pairs incorrectly rejected. These metrics, together with the decidability index, form the basis for evaluating and comparing iris recognition algorithms.

\subsection{Occlusion and Dilation Impact on Recognition}

Occlusion of the iris and pupil dilation are known factors that decrease the decidability index and negatively affect iris recognition performance. Ma et al. found that 57.7\% of false nonmatches are due to occlusion by either the eyelids or eyelashes, and additionally, that 10.7\% of false nonmatches are due to large differences in pupil sizes \cite{maEfficientIrisRecognition2004}. There is a particular difficulty with iris deformation due to pupil dilation because of the nonlinear deformation it causes on the iris \cite{clarkTheoreticalModelDescribing2013}. Wei et al. found an FNMR of 0.973\% at a FMR of 0.973\% on the CASIA IrisV3 Lamp dataset using Daugman’s model. The CASIA IrisV3 Lamp contains images of eyes with a lamp turned on and off to give variation in pupil dilation \cite{weiNonlinearIrisDeformation2007}. Hollingsworth et al. used a dataset of 18 patients to discover how eyes with different dilation levels reduced the reliability of accurate matching. This was achieved through comparing the difference of pupil-to-iris ratios (PIR). They separated the pairings into three groups of dilation levels based on their PIR: small (0.2 to 0.4), medium (0.4 to 0.6), and large (0.6 to 0.8). At a FMR of 0.1\% the large, medium, and small difference in dilation pairs had FNMR of 19.2\%, 6.4\%, and 5.9\% respectively \cite{hollingsworthPupilDilationDegrades2009}. Celik et al. further studied dilation impacts on iris matching with the compounding variable of off angle images. Using irises from 14 patients, they found average HD increased as PIR increased \cite{celikPupilDilationSynthetic2016}. Furthermore, Tomeo Reyes et al. investigated pupil dilation induced by medication using 59 patients (118 irises, 2183 total images) with PIRs of 0.265 to 0.755. They found a FNMR of 32.76\% at a FMR of 0.02\% and a FNMR of 10.10\% at a FMR of 1\% \cite{tomeo-reyesInvestigatingImpactDrug2016}. Finally, Das et al. analyzed the impact of pupil dilation on iris recognition in children aged 4 to 14 years. They found that variation in dilation within this age group accounted for 8.5\% of the variability in match score \cite{dasAnalysisDilationChildren2020}.

The largest of these studies, Wei et al., had 400 distinct eyes yielding a total of about 6,000 images for their dilated and undilated dataset \cite{weiNonlinearIrisDeformation2007}. Furthermore, the greatest spread of PIRs was seen in the work of Tomeo-Reyes et al. with a range of 0.49 (0.265 to 0.755) with 118 distinct irises \cite{tomeo-reyesInvestigatingImpactDrug2016}. At the time of this writing, we found no analysis that has been done specifically looking at eyelid occlusion and pupil dilation as compounding variables impacting iris recognition through their separate and compounding effects on HD and match rate. Intuitively, pupil dilation and eyelid manipulation are the primary sources of practical variability in the usable iris area during imaging. Therefore, examining variation in these two factors, both individually and in combination, is of particular interest in understanding their impact on iris recognition.

\begin{figure}[t]
  \centering
  \includegraphics[width=0.8\columnwidth]{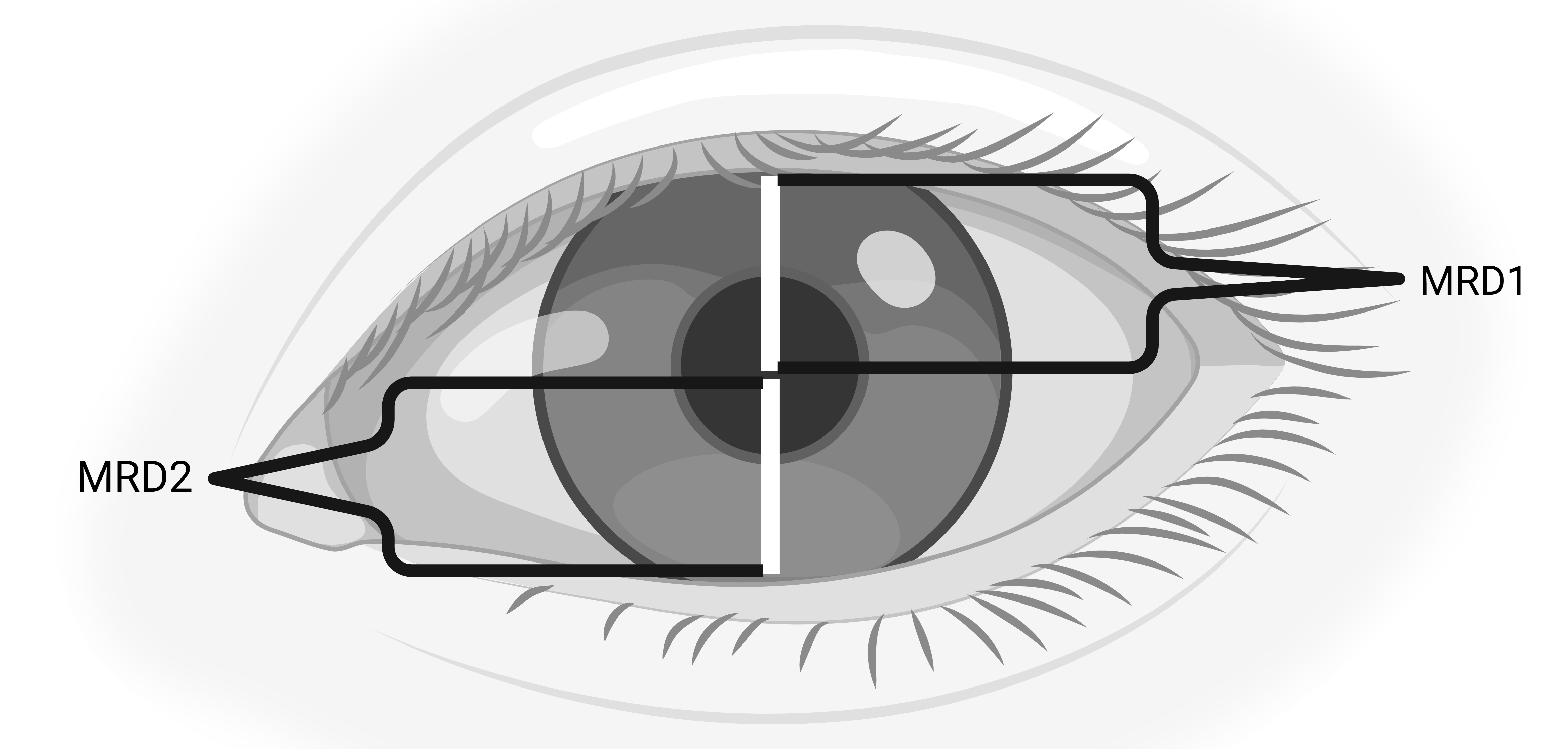} 
  \caption{Illustration of marginal reflex distances (MRD1, MRD2). MRD1 is measured vertically from the pupil center to the upper eyelid, and MRD2 from the pupil center to the lower eyelid. Distances are measured in px. Created in BioRender. Pessaud, J. (2026) https://BioRender.com/x8bi5pd}
  \label{fig:MRDFig}
\end{figure}

\section{Dataset and Software}
\subsection{Eye Enrollment}
Patients at the West Virginia University (WVU) Eye Institute were enrolled under an Institutional Review Board (IRB) approved protocol. Three sets of bilateral images were obtained for each patient using an iCAM TD100 (Iris ID Systems, Cranbury, NJ, USA):
\begin{center}
\begin{minipage}{0.75\columnwidth}
\raggedright
\begin{enumerate}
    \item Without any manipulation of the marginal reflex distance 1 (MRD1). An illustration of MRD1 is shown in Fig.~\ref{fig:MRDFig}. Classified as neutral.
    \item While patients exposed their entire iris by lifting their eyelids. Classified as wide.
    \item While patients manipulated their brow, producing partial eyelid coverage of the superior iris. Classified as squint.
\end{enumerate}
\end{minipage}
\end{center}

\begin{figure}[b]
  \centering
  \includegraphics[width=\columnwidth]{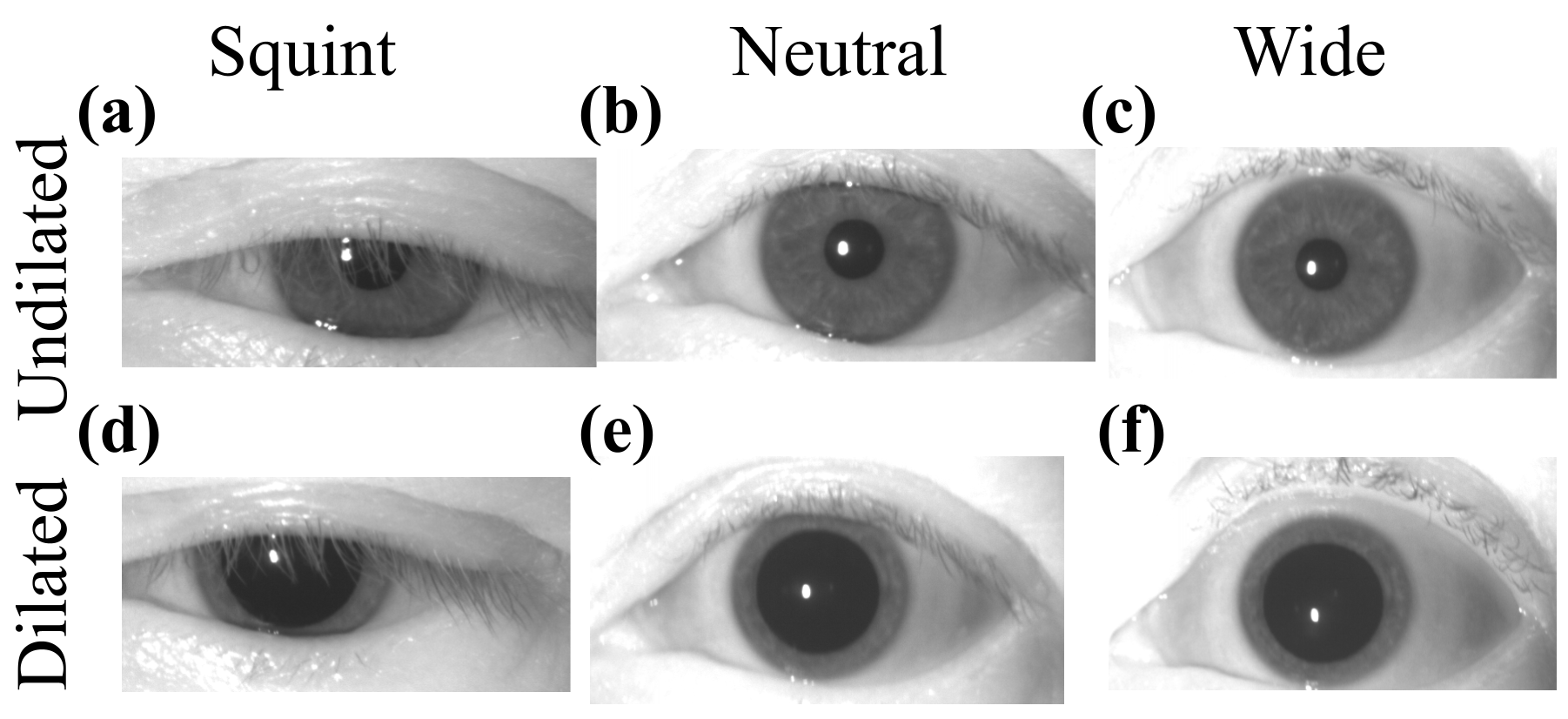} 
  \caption{Representative images for one eye across lid manipulation and dilation states. Top row: undilated (a) squint, (b) neutral, (c) wide. Bottom row: dilated (d) squint, (e) neutral, (f) wide. Images were acquired with a iCAM TD100 (Iris ID Systems, Cranbury, NJ, USA).}
  \label{EyeExFig}
\end{figure}

The patients were then dilated bilaterally using 2.5\% phenylephrine and 1\% tropicamide, and the same imaging protocol was repeated at least 30 minutes after dilation. An example of a complete set of six images for a given eye is shown in Fig. \ref{EyeExFig}.

\subsection{Recognition Software}
Worldcoin’s OpenIRIS platform was used for image segmentation and HD calculation \cite{worldcoinaiIRISIrisRecognition2023}. This open source software, released under the MIT License, processes infrared (IR) eye images to generate iris codes. Using a neural network, the pipeline classifies each pixel into one of five categories: pupil, iris, sclera, eyelash, or other. These pixel classifications are then grouped to form feature masks for the pupil, iris, eyeball, and eyelashes. The software includes multiple quality control validators to ensure sufficient image quality for reliable analysis. The most relevant of these include a PIR range of 0.1–0.7, a minimum sharpness threshold of 461 for the variance of Laplacian, two occlusion validators for 90° (maximum 25\% occluded) and 30° (maximum 30\% occluded) segments, and a minimum mask size of 4096 pixels \cite{jeanneauIrisRecognitionInference2023}.  

OpenIRIS segmentation employs a two headed architecture that combines deep learning models with traditional feature extraction methods. The eye components (iris, pupil, and sclera) and the eyelashes are segmented using separate decoding modules. The eye component decoder estimates the geometric boundaries of the eye, while the eyelash decoder identifies and suppresses noise from non-eye elements, such as eyelashes, stray hairs, or foreign material that obscure the iris texture \cite{jeanneauIrisRecognitionInference2023}. This dual segmentation framework is based on the DeepLabv3+ architecture with a MonilNet v2 backbone \cite{chenEncoderDecoderAtrousSeparable2018}.  

Following segmentation and validation, the iris is transformed from Cartesian to polar coordinates to normalize geometric variations. A Gabor filter is then applied to generate the final iris code. As described by Daugman, each grid point within the normalized iris image produces a complex value with real and imaginary components, which are encoded as two binary bits to form the unique iris code \cite{daugmanImportanceBeingRandom2003}. The pairing of iris codes is matched at the orientation that yields the lowest HD, resulting in an average HD of approximately 0.45 for two random irises, compared with 0.50 if orientation were not optimized \cite{jeanneauIrisRecognitionInference2023}.

\begin{figure}[t]
  \centering
  \includegraphics[width=\columnwidth]{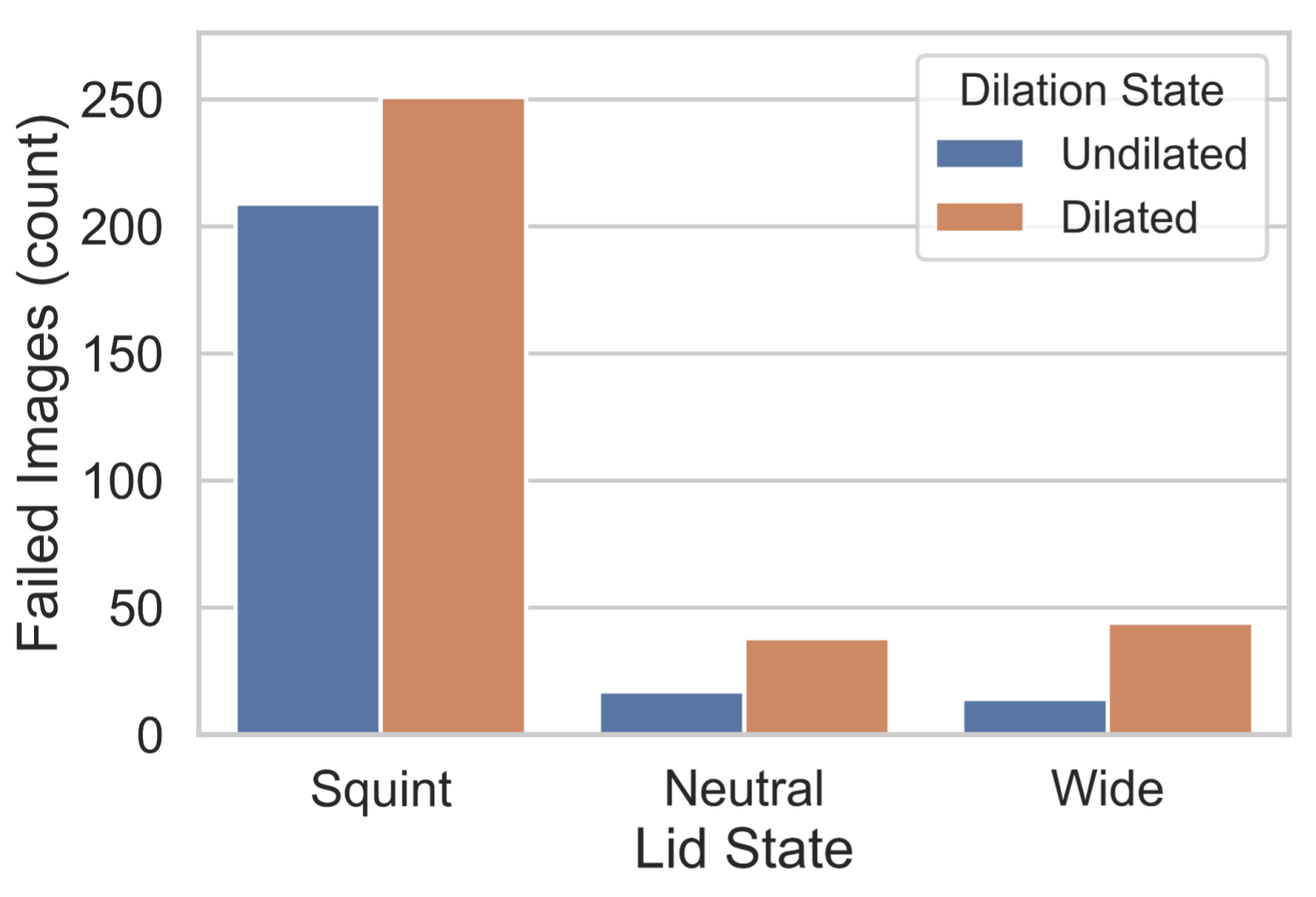} 
  \caption{Number of images that failed segmentation and iris code generation by lid manipulation and dilation state categories. Bars show the count of failed images split along lid manipulation states, squint, neutral, and wide on the x-axis; colors indicate dilation state shown in the legend. Images were failed from the OpenIRIS software with the majority due to the mask size being below 4096 px or the sharpness being below 461 (variance of Laplacian). The failure modes from pupil-to-iris ratio and occlusion were bypassed.}
  \label{FailedImgsFig}
\end{figure}

\section{Experiments and Results}
\subsection{OpenIRIS Acquisition}
A total of 278 patients (555 unique irises) were enrolled, yielding 3,341 distinct images for analysis. For this dataset, the OpenIRIS software discarded 575 (17.2\%) images due to insufficient iris code quality or segmentation failure. The breakdown of these discarded images by grouping is shown in Fig. \ref{FailedImgsFig}. Importantly, due to the nature of the investigation, the PIR range was expanded to 0.0001--0.9999, and the allowed occlusion maximums were set to 99\%. With these modifications, two major error flags were triggered during processing. The majority of images were flagged because the mask size during the encoding stage was below 4096 px, which reduced confidence in generating a reliably sized bitcode for comparison. This threshold, defined by OpenIRIS, resulted in the exclusion of 346 images. The next largest source of error, with 164 discarded images, was due to image sharpness falling below the threshold of 461.

This left a total of 2,766 usable images from 278 patients. One patient had only one eye enrolled, bringing the total number of unique irises to 555. The dataset had an overall PIR range of 0.72 (0.10--0.82) and an MRD1 range of 198.2 px (8.6--206.8 px).

\begin{figure}[t]
  \centering
  \includegraphics[width=\columnwidth]{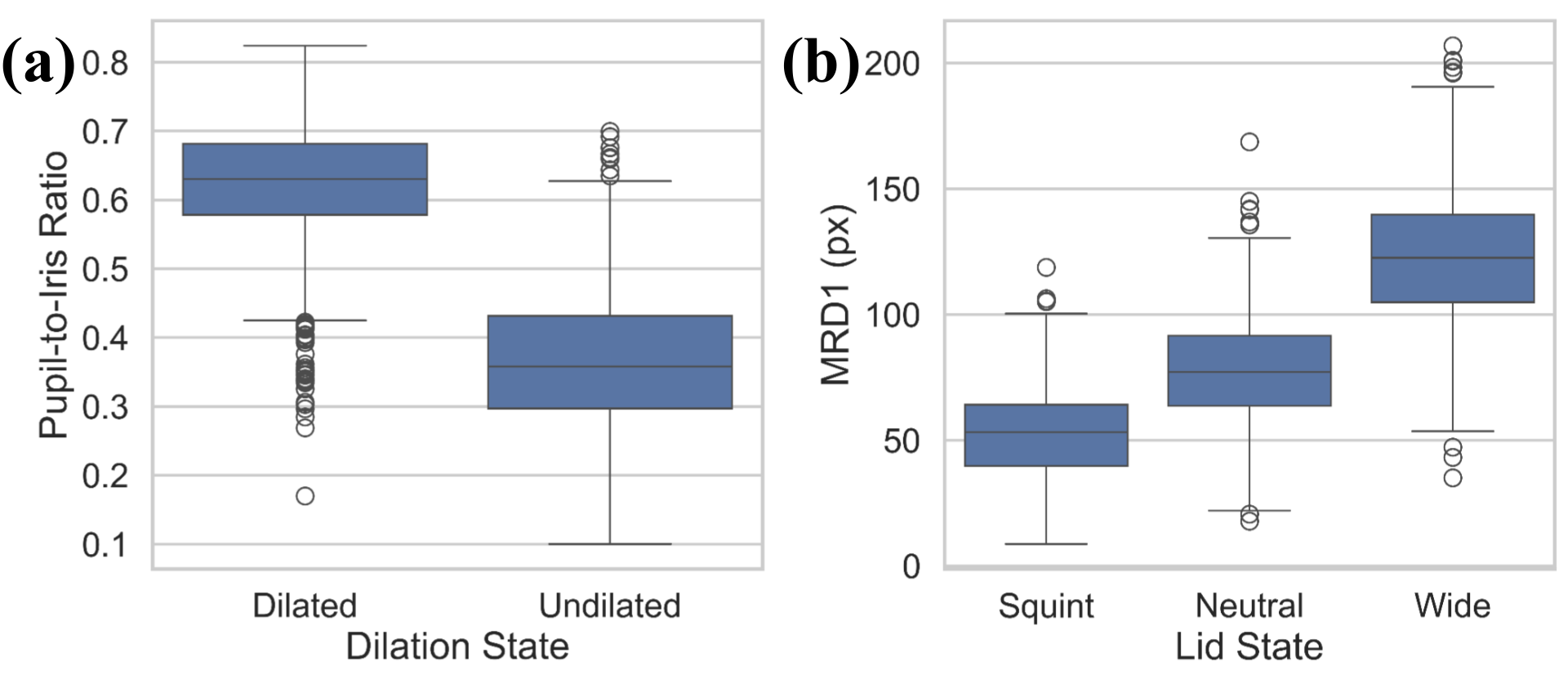} 
  \caption{Boxplots of iris metrics. (a) Distribution of pupil-to-iris ratio (PIR) separated by dilation state (dilated, undilated). (b) Distribution of marginal reflex distance 1 (MRD1) separated by eyelid manipulation state (squint, neutral, wide). Boxes show median and interquartile range (IQR) with whiskers extending 1.5xIQR. Points are beyond whiskers. PIR is unitless and MRD1 is measured in px. Dilated and undilated groups showed a significant difference (\(p << 0.001\)) in PIR using a Mann-Whitney U test. Lid state groups showed a significant difference (\(p << 0.001\)) in MRD1 with a Kruskal-Wallis test.}
  \label{fig:two-up}
\end{figure}

\subsection{Metrics}
Numerous features were calculated for each image including: PIR, MRD1, and visible iris area (VIA). OpenIRIS estimates the diameters of the iris and pupil by finding the maximum distance of any two points on the polygon of the respective structure (iris or pupil). The PIR is then calculated by dividing the pupil diameter by the iris diameter. MRD1 was calculated by finding the vertical distance between the center of the pupil and where the eyeball mask ends at its top. Clinically, MRD1 is measured as the distance from the corneal light reflex to the upper eyelid margin; however, for practical purposes, the center of the pupil has been used instead. MRD1 is measured in pixels. The VIA is calculated as the sum of the iris mask pixels after subtracting the pupil mask.

\begin{figure}
  \centering
  \includegraphics[width=\columnwidth]{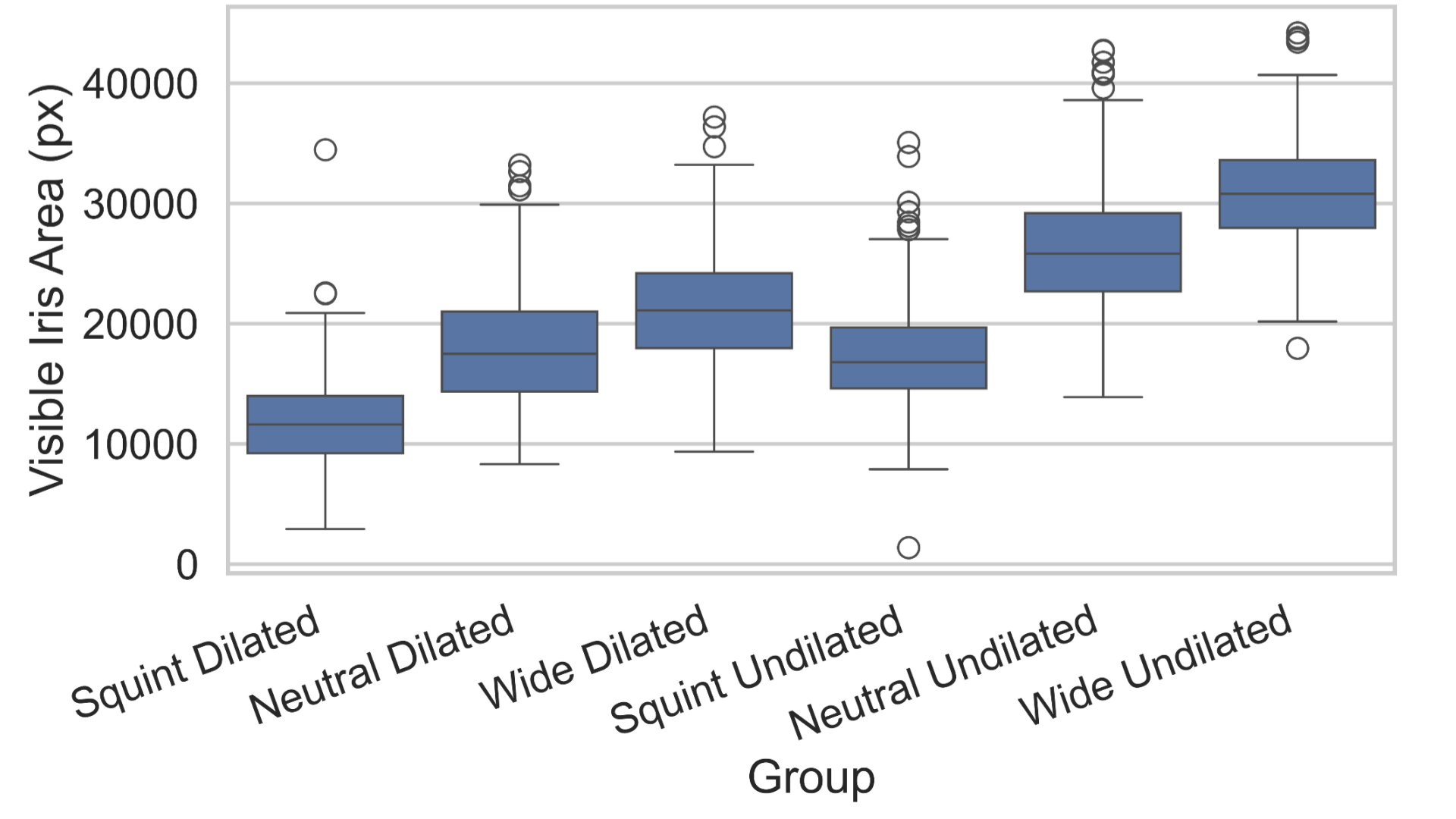} 
  \caption{Box plots of the distribution of visible iris area (VIA) split by combinations of lid state (squint, neutral, wide) and dilation state (dilated, undilated). Boxes show median and interquartile range (IQR) with whiskers extending 1.5xIQR. Points are beyond whiskers. VIA is measured in px. There is a significant difference (\(p << 0.001\)) of VIA between the six groups with a Kruskal-Wallis test.}
  \label{VIAFig}
\end{figure}

Categorically, the data acquisition labeling was used to classify images with each iris labeled as squint, neutral, or wide, and as either dilated or undilated. Lid state and dilation were analyzed quantitatively using MRD1 and PIR, respectively. These attributes are intuitively related, as shown in Fig. \ref{fig:two-up}. The combined effects of lid state and dilation were further analyzed quantitatively through VIA. Fig. \ref{VIAFig} illustrates how all six categories exhibit distinct VIA distributions.

\subsection{Pairing}
Eyes were paired into two groups: matching pairs (i.e., two distinct images of the same eye) and nonmatching pairs (i.e., images of two different eyes). The threshold used for HD was 0.38, as recommended by OpenIRIS. All possible pairings of eyes were analyzed. The matching eyes produced 5,834 total pairs, while the nonmatching eyes produced 3,818,161 pairs, yielding a total of 3,823,995 pairs.  

\begin{figure}[t]
  \centering
  \includegraphics[width=\columnwidth]{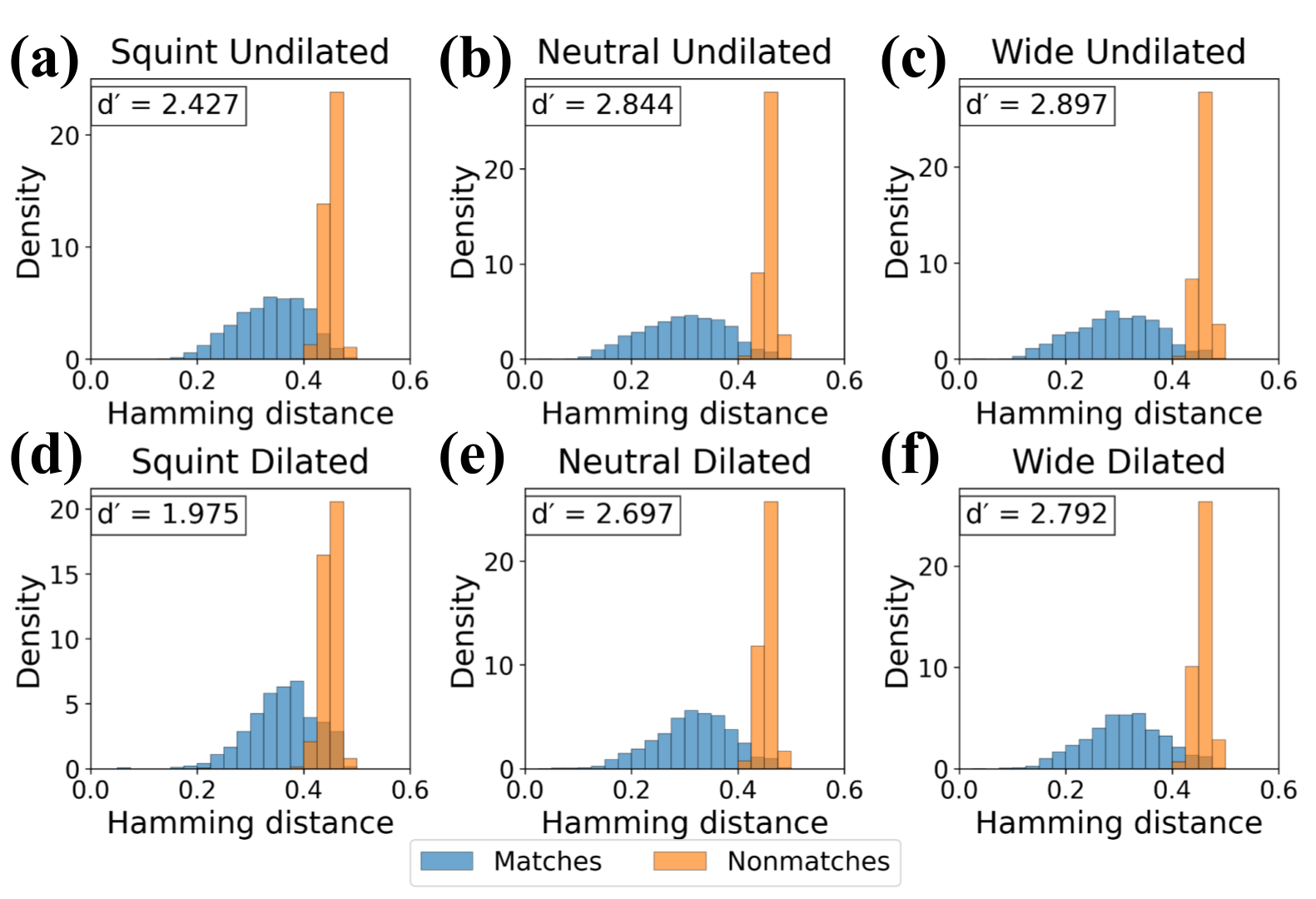} 
  \caption{Decision environments for each dataset relative to a single enrollment image split by lid state (squint, neutral, wide) and dilation state (undilated, dilated). Top row: undilated (a) squint, (b) neutral, (c) wide. Bottom row: dilated (d) squint, (e) neutral, (f) wide. Histograms show Hamming distance (HD) distributions for matching and nonmatching pairs, shown in the legend (bin width = 0.125 HD). A random sampling of nonmatching pairs was plotted equal in number to the set of matching pairs for that group. The decidability index summarizes group separation, $d' = \frac{|\mu_1 - \mu_2|}{\sqrt{0.5(\sigma_1^2 + \sigma_2^2)}}$.}
  \label{DecEnvFig}
\end{figure}

The eyes in a given pair were classified as an enrollment image and a probe image. The enrollment image represents the iris code that would be stored in a database for personal identification, while the probe image represents the image captured for comparison against the enrolled iris code. To simulate practical applications of iris recognition, the enrollment image was held categorically constant, while the probe image varied in lid state and dilation. A wide, undilated eye was selected as the enrollment image for all pairings. This ensured that each pair included at least one undilated eye with a wide lid state. A wide, undilated enrollment image was chosen because those pairings yielded the highest decidability index, as shown in Fig. \ref{DecEnvFig}. Pairs containing this enrollment type had the largest decidability index of 2.897. A neutral, undilated enrollment had a decidability index of 2.844, and a wide, dilated enrollment had an index of 2.792. All other enrollments had decidability indices below 2.7, with the smallest at $d' = 1.975$ for a squint, dilated enrollment. If both eyes in a pair were undilated and wide, the enrollment and probe labels were assigned at random.

\begin{figure}[b]
  \centering
  \includegraphics[width=\columnwidth]{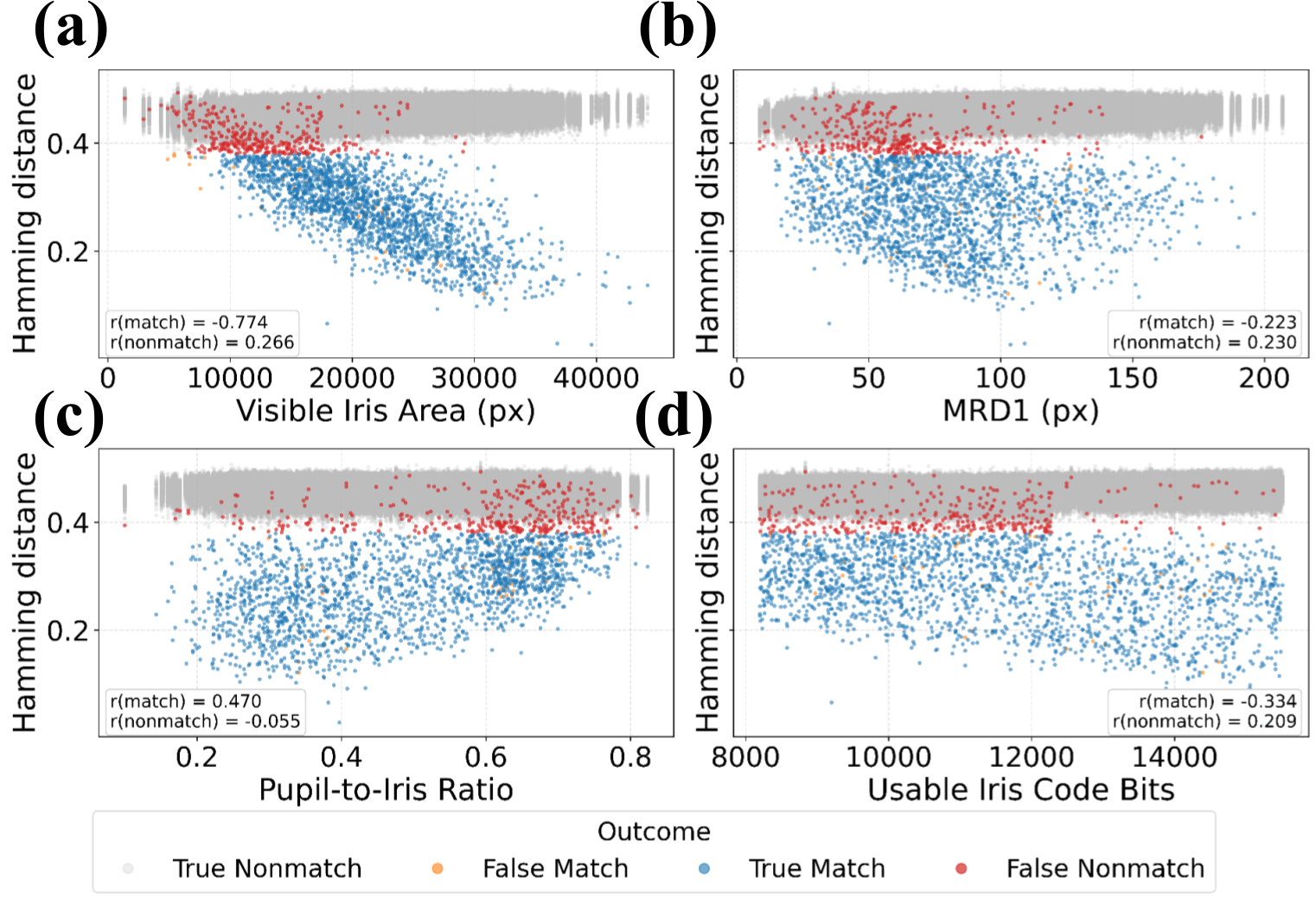} 
  \caption{Scatter plots of probe image features vs. Hamming distance (HD) to a wide, undilated enrollment eye. (a) Visible iris area (VIA) shows the strongest negative correlation with HD ($r=-0.774$). (b) MRD1 has a weak negative correlation ($r=-0.223$). (c) Pupil-to-iris ratio (PIR) shows a moderate positive correlation ($r=0.470$). (d) Iris code length shows a weak negative correlation ($r=-0.334$). Points are color-coded by pairing type, with Pearson’s $r$ reported separately for matches and nonmatches.}
  \label{WUFig}
\end{figure}

\subsection{Hamming Distance Correlations}
Quantitative analysis was performed using VIA, PIR, and MRD1 values from the probe image in each pair. All correlations are reported as Pearson correlation coefficients. As shown in Fig. \ref{WUFig}, VIA exhibited the strongest correlation with HD ($r = -0.774$). PIR and MRD1 showed weaker correlations of $r = 0.470$ and $r = -0.223$, respectively. All variables in nonmatching pairs had correlations of \(|r| < 0.27\), indicating no appreciable relationship between these features and HD for nonmatching pairs.  

Additionally, iris code length was evaluated. Although it had moderate correlation with VIA ($r = 0.501$), it did not show a strong correlation with HD ($r = -0.334$). Image quality metrics such as contrast, sharpness, pupil circularity deviation, distinguishability of the iris boundaries between the pupil and sclera, iris and pupil diameters, and grayscale utilization were computed using the BIQT toolkit developed by MITRE \cite{BIQTIris2023}. This open source software is licensed under the Apache 2.0 License. All BIQT features exhibited weak correlations with HD (\(|r| < 0.5\)) and were therefore not analyzed further.

\begin{figure}[t]
  \centering
  \includegraphics[width=\columnwidth]{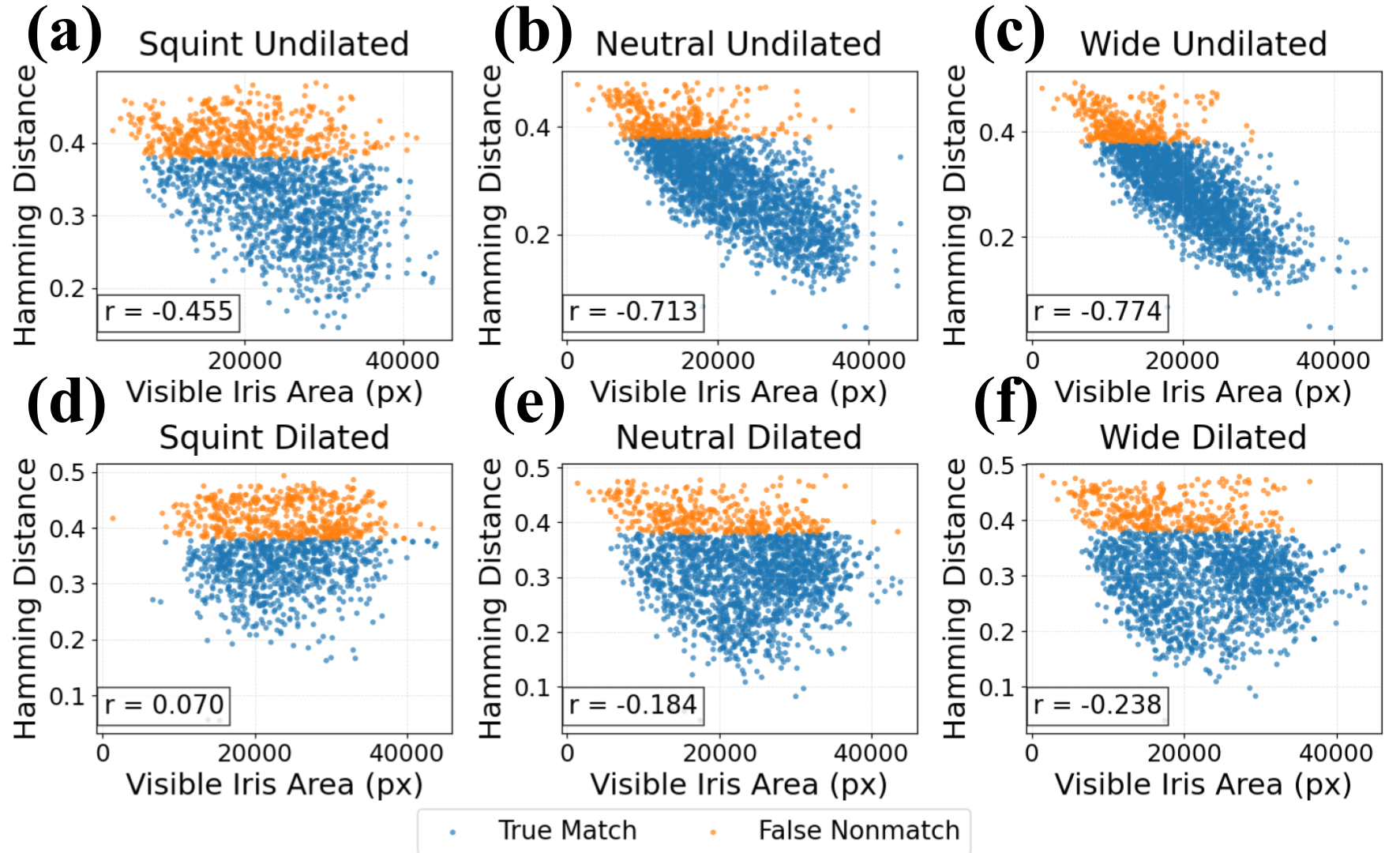} 
  \caption{Scatter plots of visible iris area (VIA) of the probe image versus Hamming distance (HD), separated by enrollment image condition. Top row: undilated (a) squint, (b) neutral, (c) wide. Bottom row: dilated (d) squint, (e) neutral, (f) wide. Data points are color-coded by match type (true match vs. false nonmatch). The strength of correlation between VIA and HD varies with enrollment condition, with the strongest correlation observed for wide, undilated reference images ($r=-0.774$).}
  \label{VIAEnrFig}
\end{figure}

Another notable effect observed in the VIA correlation was its variation with the enrollment image condition. As shown in Fig. \ref{VIAEnrFig}, iterating through all possible enrollment image types revealed that the correlation with HD decreased as MRD1 became smaller or when a dilated pupil was used for the enrollment. There was a measurable drop from the wide, undilated reference to the neutral, undilated reference, with HD correlation changing from $r = -0.774$ to $r = -0.713$. The effects of eyelid manipulation did not appear to influence the correlation as strongly as pupil dilation, with all dilated enrollment references exhibiting reduced correlations compared to undilated ones. The dilated references followed a similar decreasing trend across eyelid conditions as observed in the undilated group.

\subsection{Quality Gate Modeling}
A logistic regression model was developed to evaluate the practical impact of the observed correlations during image acquisition. Using the same dataset of wide, undilated eyes as the enrollment images, the model generated a quality gate predicting whether a given probe image would be correctly classified as a match or nonmatch based on its measured features.  

For each probe image, logistic regression estimated the probability that iris recognition would correctly classify the pair as a match. Since each probe image appeared in multiple pairings, the probabilities from all its pairings were averaged to produce a single quality score for that image. The model then computed the FNMR under two constraints. First, the HD threshold was fixed to ensure the FMR remained below 0.1\%. Second, the quality gate threshold was varied to evaluate FNMR across discard rates ranging from 0--7\%. The gate was applied by removing probe images in the lowest percentiles of quality scores (e.g., a 5\% discard rate excluded the bottom 5\%). All models were bootstrapped with 500 resamples to generate 95\% confidence intervals.  

\begin{figure}[t]
  \centering
  \includegraphics[width=\columnwidth]{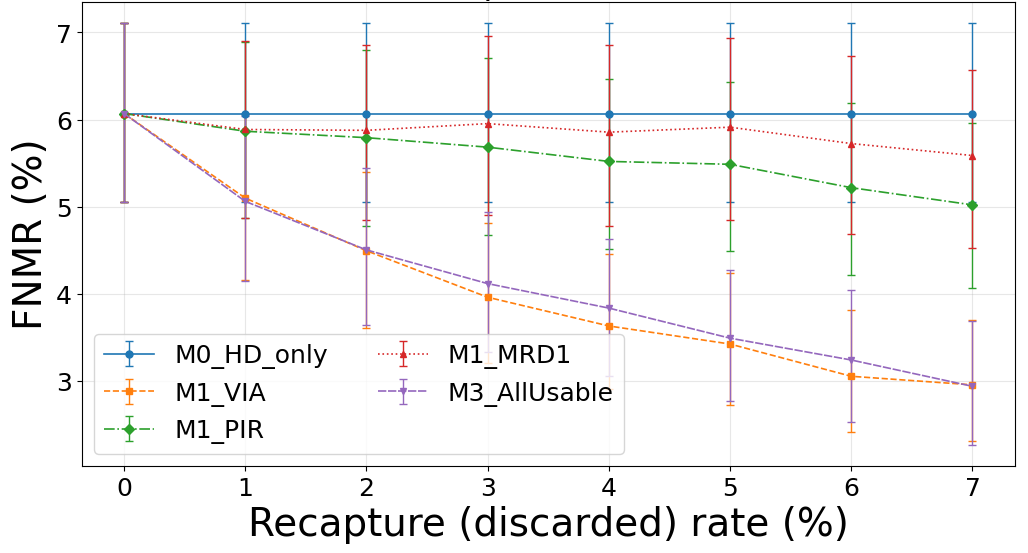} 
  \caption{False nonmatch rate (FNMR) as a function of recapture rate under different modeling parameters. M0 denotes no quality control. Single-feature models are M1\_VIA (visible iris area), M1\_PIR (pupil-to-iris ratio), and M1\_MRD1 (marginal reflex distance 1), while M3 represents the combined model using all three features. Logistic regression was used to assign a quality score to each probe image, defined as the average predicted probability of correct classification across all its pairings. FNMR was evaluated under two constraints: (i) the HD threshold was fixed such that the false match rate (FMR) was below 0.1\%, and (ii) the quality gate threshold was varied to discard the bottom 0--7\% of probe images by quality score. All models were bootstrapped with 500 resamples to generate 95\% confidence intervals.}
  \label{ModelFig}
\end{figure}

Fig. \ref{ModelFig} shows the relationship between FNMR and discard rate. The control model (M0) represents the baseline using HD alone, without any quality gating. The single-feature models (M1) included PIR (M1\_PIR), MRD1 (M1\_MRD1), and VIA (M1\_VIA), while the combined model (M3) incorporated all three features simultaneously.  

All models achieved mean FMR values between 0.07\% and 0.10\%. As shown in Fig. \ref{ModelFig}, VIA was the most effective predictor of image quality, yielding consistently lower FNMR values across all discard rates compared with PIR or MRD1 alone. Furthermore, the combined model (M3) performed nearly identically to the VIA-only model, indicating that PIR and MRD1 contributed minimal additional predictive value beyond VIA.

\section{Discussion}
This dataset highlights the distinct advantages of VIA as a single, robust feature for assessing probe image quality in iris recognition. Among all evaluated metrics, VIA demonstrated the strongest correlation with HD in matching pairs and no appreciable correlation among nonmatching pairs, making it uniquely suited for identifying recognition reliability.  

A key practical strength of VIA is that it can be computed directly from the probe image alone, without requiring comparison to an enrollment image or iris code. This means quality assessment can occur immediately after segmentation, well before any matching or encoding step. Such independence from enrollment images significantly improves efficiency, enabling real-time feedback during image capture and avoiding the computational burden of pairwise comparisons. In contrast, other metrics such as PIR or MRD1 require both pupil and eyelid geometry to be considered relative to each other and provide weaker predictive power.  

The VIA–HD relationship was most pronounced when the enrollment image was captured in a wide, undilated state, yielding a correlation of $r = -0.774$ and the highest decisional separability ($d' = 2.897$). A neutral, undilated enrollment also performed well ($d' = 2.844$), suggesting that undilated eyes in general preserve most of the VIA advantage. The reduction in correlation under dilated conditions indicates that pupil dilation likely alters usable iris area geometry more substantially than lid position. Thus, while wide, undilated enrollments provide optimal reliability, VIA likely remains effective under neutral, undilated conditions; however, this interpretation should be considered in light of the exclusion of many squint images during analysis.

Importantly, VIA showed minimal correlation in nonmatching pairs ($r = 0.266$), confirming that variability in VIA does not affect the ability of the system to reject incorrect matches. This distinction underscores that VIA primarily improves FNMR without negatively influencing FMR, which is an essential property for practical deployment.  

In contrast, iris code length exhibited a weaker relationship with HD and offered limited utility as a quality indicator. Although code length correlated moderately with VIA ($r = 0.501$), it did not translate into a significant correlation with HD ($r = -0.334$) or a meaningful improvement in recognition accuracy. This suggests that longer or shorter iris codes, in themselves, do not necessarily reflect usable image quality.

We also examined whether the difference in VIA between probe and enrollment images provided a stronger predictive relationship with HD. This differential metric showed a lower correlation ($r = 0.668$) compared to the probe image VIA alone (\(|r| = 0.774\)). This finding highlights a practical advantage of VIA, as it can be computed independently from the probe image without requiring access to the enrollment image or iris code. As a result, VIA-based quality control can occur immediately after segmentation, eliminating the need for post-processing or pairwise comparison and enabling faster, more efficient image assessment during acquisition.

The VIA-only logistic regression model further supports these findings (Table \ref{tab:via-discard-fmr-fnmr}). Discarding as little as 3--5\% of low-quality probe images based on VIA reduced FNMR by more than 35\% while maintaining or slightly improving FMR. Since VIA relies solely on segmentation, this process can occur nearly instantaneously after capture, providing operators with immediate feedback on image acceptability.  

In practice, three considerations are relevant for deploying a quality gate approach using probe image VIA. First, segmentation must be fast and reliable to enable real-time VIA computation.  Second, there must be an allowance of repeated imaging for a certain percentage of probed irises. There may be some practical concerns with having discard rates as high as 7\% as shown in the quality gate modeling. However, it is not expected for the probe images to have the large variability seen in this dataset in more controlled environments for image acquisition. With proper lighting and instructions for wide lid manipulation, the increased confidence could be gained with a smaller discard rate. Third, the observed effects predominantly influence FNMR, indicating a stronger imperative to reduce FNMR relative to FMR. Nonetheless, since FMR also decreases, this improvement does not represent a compromise of FMR for enhanced FNMR performance, but rather that FNMR exhibits a comparatively greater reduction.

The relative importance of VIA will ultimately depend on the use case of the iris recognition system. Applications that prioritize verification accuracy, such as border control or secure access, may benefit most from VIA’s ability to reduce FNMR without increasing FMR. In contrast, high-throughput identification systems may balance stricter quality gating against operational efficiency. This flexibility underscores the practical relevance of VIA as a tunable, context-dependent quality feature rather than a fixed performance constraint.

\begin{table}[t]
  \centering
  \caption{Average false match rate (FMR) and false nonmatch rate (FNMR) at different discard rates using visible iris area (VIA) only logistic regression modeling.}
  \label{tab:via-discard-fmr-fnmr}
  \begin{tabular}{lcc}
    \hline
    \textbf{Discard Rate (\%)} & \textbf{Mean FMR (\%)} & \textbf{Mean FNMR (\%)} \\
    \hline
    0 & 0.10 & 6.07 \\
    1 & 0.09 & 5.10 \\
    2 & 0.09 & 4.50 \\
    3 & 0.08 & 3.96 \\
    4 & 0.08 & 3.63 \\
    5 & 0.08 & 3.43 \\
    6 & 0.08 & 3.06 \\
    7 & 0.07 & 2.96 \\
    \hline
  \end{tabular}
\end{table}

Taken together, these results demonstrate that VIA is a powerful, independent indicator of probe image quality that can be implemented efficiently within existing iris recognition pipelines. Its ability to operate without enrollment comparison, combined with strong predictive value and real-time feasibility, positions VIA as a practical and scalable feature for improving recognition reliability in large-scale biometric systems.

\section{Conclusions and Future Work}
This study quantitatively analyzed how lid state and pupil dilation impact iris recognition, both individually and in combination. Using a comparatively large dataset with high variability introduced by controlled lid manipulation and dilation drops, we identified a strong correlation between VIA and recognition performance. The strong relationship with HD indicates that VIA may be a better predictor of matching accuracy compared to the more commonly used PIR. Additionally, the modeling demonstrates a strong potential use case for VIA as a simple quality control metric prior to iris code calculation.  

These findings suggest that analyzing VIA during iris image acquisition may improve confidence in matching under high-variability conditions. For instance, this approach may be beneficial in environments with variable lighting that alters pupil size, among older populations with a higher prevalence of lower MRD1 values, or in patients with iris anomalies such as traumatic mydriasis. Future work should further explore models incorporating VIA into iris acquisition and recognition pipelines.

\ifCLASSOPTIONcaptionsoff
  \newpage
\fi
\bibliographystyle{IEEEtran}
\bibliography{bibtex/bib/IradDraft}

\end{document}